\let\clineorig\cline
\documentclass[referee,pdflatex,sn-basic, iicol]{sn-jnl}
\let\cline\clineorig

\usepackage{amsmath,amsfonts}%
\usepackage{algorithm}%
\usepackage{array}%
\usepackage[caption=false,font=normalsize,labelfont=sf,textfont=sf]{subfig}%
\usepackage{textcomp}%
\usepackage{stfloats}%
\usepackage{url}%
\usepackage{verbatim}%
\usepackage{graphicx}%
\usepackage{hyperref}%
\usepackage{multirow}%
\usepackage{natbib}%

\usepackage[justification=centering]{caption}%
\graphicspath{ {./} }%
\begin{document}

\title[Article Title]{HilEnT: \textbf{Hil}bert, \textbf{En}tropy \textbf{T}ransformed Image Based Malware Detection}

\author*{\fnm{Rahul} \sur{Kale}}\email{rahulvishwanath.kale@stengg.com}

\author{\fnm{Thesath} \sur{Wijayasiri}}\email{boperachchigethesathguwantha.wijayasiri@stengg.com}

\author{\fnm{Kar Wai} \sur{Fok}}\email{fok.karwai@stengg.com}

\author{\fnm{Vrizlynn L. L.} \sur{Thing}}\email{vriz@ieee.org}

\affil{\orgdiv{Cybersecurity Strategic Technology Centre}, \orgname{ST Engineering}, \orgaddress{\country{Singapore}}}

\abstract{With the increasing threat of malware across various software related domains, malware detection and classification is critical to determine the response actions.
	Different strategies have been adopted to address the challenge of malware detection. With the advent of deep learning techniques, malware detection using image processing has garnered research attention.
	In this work, we proposed a novel malware binary to image transformation technique \textit{HilEnT} based on a combination of Hilbert curve based transformation of malware binary and the entropy feature comparison of malware file with benign and malware classes.
	Three grayscale images produced during this process are combined to form a three-channel colored image which is then used for malware detection using machine learning techniques.
	We performed supervised binary and multiclass classification to evaluate the effectiveness of our proposed HilEnT.
	We also evaluated a few-shot learning technique to assess the robustness of our proposed HilEnT in a practical setting where the number of available class samples is limited.
	Furthermore, we investigated the benefits of combination of Histogram of Oriented Gradients and Principal Component Analysis for time performance improvements through feature reduction techniques.
	We evaluated our proposed methodology on four datasets: Dike, Michael Lester Dataset, MicrosoftBIG 2015 and a self-collected dataset, and achieved the state-of-the-art results.}

\keywords{Malware Detection, Machine Learning, Malware Classification, Neural Network}

\maketitle

\section{Introduction}\label{sec1}
Malicious Software (Malware) has been an ever-growing problem in software-dependent industries as they can potentially infect various computing and networking devices. According to a survey \citep{bensaoud2024survey} published recently in 2024, the number of new vulnerabilities is still increasing every month in the year 2024. As the malware detection systems are developed based on the existing available malware, new malware designed by attackers may evade detection on such detection systems. With such a large number of large monthly additions, distinguishing and classifying different types of malware is crucial to determine the response actions for safeguarding the systems under attack. This process also assists in identifying their potential impacts on the system and selecting a defense mechanism for protection against them.

Different malware detection approaches \citep{nataraj2010detecting,bensaoud2024cnn,natani2013malware,chuang2015machine,li2024syndroid,wang2023dae} have been explored in the literature.
In practical applications, most of the network traffic is generally benign. With the constantly evolving nature of the malware, it is generally difficult to obtain large number of malware samples of each class for thorough training of supervised malware classification algorithms. However, obtaining a few labeled anomaly samples is comparatively less expensive and more practical. With few-shot detection approach, the limited labeled data is leveraged for effective malware classification. This can be achieved either by fine-tuning the pre-existing models with handful of labeled data or by performing data augmentation of existing data. Additionally, these models can potentially be used for detecting unknown classes not used/seen during training. Some recent works \citep{hsiao2019malware,bai2020unsuccessful,wang2021novel, conti} explore few-shot learning based malware detection.

Converting malware binary features to images, also known as malware visualization \citep{nataraj2011malware}, is one of the popular machine-learning (ML) approaches adopted recently for malware detection and classification\citep{nataraj2011malware, cui2018detection,hemalatha2021efficient,lo2019xception,luo2017binary, roseline2020intelligent,vinayakumar2019robust, vu2020hit4mal}.
The key idea here is to tackle the malware detection problem as an image processing problem with Convolutional Neural Networks (CNNs) being commonly used for such applications. In this paper, we adopted this approach of converting the file's binary sequence into an image which is then analyzed using machine learning techniques for malware detection and classification. 
We proposed a novel malware binary to image transformation technique \textit{HilEnT} based on a combination of Hilbert curve transformation, benign class entropy cutoff comparison and malware class entropy cutoff comparison.
Hilbert curve transformation converts the malware binary into grayscale Hilbert curve pattern based image, while benign class entropy and malware class entropy based cutoff comparison provide one image each containing corresponding regions of interests for both classes.
These three grayscale images are combined to form a three-channel colored image which is then used for malware detection using machine learning techniques. We performed supervised binary and multiclass classification to evaluate the effectiveness of our proposed HilEnT.
Furthermore, we investigated the benefits of combination of Histogram of Oriented Gradients (HOG) and Principal Component Analysis (PCA) for time performance improvements through feature extraction and dimensionality reduction techniques.
We also evaluated few-shot learning technique to assess the robustness of our proposed HilEnT in a practical setting where number of available class samples are limited.

To summarize, in this work we first proposed a novel binary file to image transformation technique for malware visualization, called \textit{HilEnT} based on Hilbert curve and entropy approaches. We then evaluated its effectiveness using a supervised learning based CNN and robustness using a few-shot learning based technique. The performance evaluation was conducted using four datasets: Dike, Michael Lester Portable Executable(PE) Dataset, MicrosoftBIG 2015 and a self-collected dataset.

The rest of the paper is organized as follows: In Section \ref{sec:related}, we review recent related works in literature about malware detection approaches including malware visualization and few-shot learning based methods along with some general approaches for malware detection. The proposed HilEnT framework is described in Section \ref{sec:framework}. In Section \ref{sec:setup}, the details about the dataset and experimental setup are discussed. The evaluation experiments and their results are presented in Section \ref{sec:results}.
Finally, the paper is concluded in Section \ref{sec:conclusion}.

\section{Related Works}\label{sec:related}

\subsection{General Machine Learning-based approaches for Malware Detection}

\citet{nataraj2010detecting} utilized bigram-based features and support vector machines to distinguish packed and unpacked executables. The classification was performed on the raw binary data with a focus on the fast processing time.
For malware classification, \citet{bensaoud2024cnn} first extracted opcode sequences and Application Programming Interface (API) calls from Windows malware files and then transformed them into N-gram sequences. CNN and Long Short-Term Memory was then utilized for classification task.
\citet{natani2013malware} identified that various malware behaviours are associated with the functions utilized by compromised files through system library calls. Therefore, they used API function frequency as feature vector with ensemble classifiers to perform malware classification.
\citet{chuang2015machine} utilized API calls for Android malware analysis. A Support Vector Machine (SVM)-based hybrid model was proposed which was trained using separate malware-preferred feature set and benign-preferred feature set.
For Android malware detection and classification, SynDroid model was proposed by \citet{li2024syndroid}. This model was designed to address class imbalance by first generating high-dimensional samples using CTGAN-SVM and then utilize Random Forest for malware classification.
\citet{wang2023dae} first preprocessed the malware, and subsequently HOG of the grayscale malware image was obtained followed by the autoencoder for dimensionality reduction. Finally an ensemble of Extra Trees, XGBoost and Random Forest was used with voting for final classification decision. The dataset used for this work was Microsoft's 2015 Malware Classification Challenge. During malware preprocessing, the malware binary file is converted to grayscale using B2M algorithm, and image scaling is performed using nearest-neighbour interpolation algorithm. 
The main purpose of the deep autoencoder is for dimensionality reduction as HOG may generate redundant features due to overlapping blocks.
A related hybrid architecture was proposed by \citet{kumar2024cnn}, who combined convolutional feature extraction with an autoencoder to improve classification performance for image-transformed malware in Industrial IoT settings. This further highlights the growing trend of deep models integrating dimensionality reduction modules for malware analysis.
Though our proposed approach evaluates HOG as a first step similar to this existing work, our approach uses PCA as a second step for dimensionality reduction. This existing work uses autoencoder for dimensionality reduction. PCA may provide performance benefit for dimensionality reduction. Our proposed work converts the malware binary features into 3-channel images instead of a single channel grayscale images. We also utilize few-shot classification method using CNN implementation for malware classification whereas \citet{wang2023dae} utilizes the ensemble learning based detection approach.

\subsection{Malware Detection with Malware Visualization}

\citet{nataraj2011malware} were the first to propose the idea of malware visualization using bytes of a malware file to represent pixels within an image. CNN was used for classification of the grayscale images obtained by transforming the malware binary files.
\citet{cui2018detection} utilized deep learning for malicious code detection using malware visualization. The malware binary bit strings were converted to grayscale images and Bat algorithm was used for data augmentation to address the class imbalance. Finally, CNN was utilized for malware classification.
\citet{hemalatha2021efficient} used deep learning models for addressing data imbalance and achieve high accuracy for malware classification on four datasets. Malware binaries were transformed into two-dimensional images and DenseNet was utilized for the classification.
\citet{lo2019xception} proposed a deep learning method for classification of grayscale malware images obtained by transforming malware binaries. The deep learning model utilized was based on the CNN with Xception model which reduces the overfitting issues. Their proposed model achieved high accuracy for image based malware classification and outperformed other ML methods.
\citet{luo2017binary} proposed a 3-step method for malware classification. In the first step, malware binary was transformed into a grayscale malware image and then reorganized in a $3 \times 3$ grid. In the second step, local binary pattern (LBP) features were extracted through these reorganized grids. In the final step, a CNN was utilized for malware classification.
\citet{roseline2020intelligent} proposed a layered ensemble approach for malware detection and classification based on malware visualization. This approach was designed to reduce model complexity by removing the need of hyperparameter tuning or backpropagation. After transforming the malware binaries to grayscale images, sliding window scanning and cascade layering was used for malware classification.
\citet{vinayakumar2019robust} proposed a novel image processing technique specifically tuned for deep learning methods to obtain a robust zero-day malware detection model. In their proposed approach, malware binary was converted to a grayscale image first. Then the flattened grayscale image array was passed through 1-dimensional convolutional and pooling layers followed by LSTM and fully connected layers for malware classification.
The method proposed by \citet{vu2020hit4mal} focused on pixel encoding and byte arrangement for malware visualization of the malware binaries. Space-filling curves were used to enhance these images with statistical features which assisted in improved detection of malware.
More recently, \citet{andriani2025cnnautomic} proposed CNN-AutoMIC, which also uses Nataraj style grayscale malware images as input but focuses on a robust two-stage architecture combining a CNN feature extractor, a nonlinear autoencoder that compresses the features into a 2D latent space, and a KNN classification engine. Their emphasis is on cross dataset robustness and explainability rather than proposing a new visualization method.
IMCFN, introduced by \citet{vasan2020imcfn}, fine-tunes a deep CNN backbone for malware images and remains one of the strongest recent CNN-only visualization-based baselines, showing that convolutional models trained on bytecode images can achieve competitive performance.
Similarly, \citet{ashawa2024enhanced} demonstrated that enhanced CNN architectures continue to be highly effective for image-based malware detection, further reinforcing the relevance of visual feature extraction pipelines.
This validates that deep CNN models tend to show remarkable results when combined with malware visualization.
Transfer learning approaches have also gained attention, with \citet{panda2023transfer} showing that IoT malware images can be classified effectively using pretrained CNN backbones.
\subsection{Malware Detection with Few-shot Learning}
\citet{fei2006one} were the first to propose the idea of few-shot learning. \citet{hsiao2019malware} adopted Siamese neural networks for malware image classification. During the preprocessing step, grayscale images were obtained from malware binaries and subsequently classified using average hash. One-shot learning using Siamese networks was performed for malware classification.
For Android malware classification, \citet{bai2020unsuccessful} used few-shot learning  by adopting Siamese neural network. Multi-layer Perceptron network was trained for transforming malware into a latent representation in a continuous vector space using Siamese neural networks which improved classification performance.
\citet{wang2021novel} proposed `SIMPLE', which was a few-shot malware classification model and utilized multi-prototype modeling. It was based on their observation of multimodal data distribution for the behaviors of malware within same malware family. Their proposed model achieved high accuracy for 5-way 5-shot tasks.
\citet{conti} proposed an approach which combined malware visualization technique and few-shot learning approach. The malware feature visualization technique generated a 3-channel image by fusing Entropy Image, Markov Image and Gray-level Images obtained from malware binary. The 3-channel fused image was then used with few-shot classifiers. Authors proposed two few-shot classification methods, Convolutional Siamese Neural Network(CSNN) and baseline Feature Extractor(baselineFE). They performed the evaluation experiment for one-shot and ten-shot support sets which indicates the number of training samples available for each class. CSNN approach for one-shot classification was utilized for the case with limited amount of training data. For 10-shot classification, the baselineFE approach was utilized which outperformed the CSNN approach but required more training data. Additionally, both methods were tested for novel/unseen class cases. Authors evaluated their proposed method on three datasets: Malimg, Microsoft BIG 2015 and Malbaz. The first two datasets are publicly available whereas the third dataset was compiled using a dataset consisting of recently published malware executables on a public malware repository MalwareBazaar. 
More recent work such as \citet{alfarsi2024fewshot} has reaffirmed the usefulness of few-shot learning for malware classification, particularly in scenarios where labelled samples per family are scarce.
Meta learning approaches have also emerged, including Mi-MAML by \citet{ma2024mimaml}, which adapts few-shot optimisation strategies to handle rapidly evolving malware families.
Our proposed HilEnT malware visualization technique based on grayscale Hilbert curve transformation, as well as entropy cutoff comparison approach differs from approach in \citet{conti}.
In their work, Markov image is generated by using the bigrams within the binary file. Each pixel intensity value in the resultant image represents frequency of value of the bigram represented by the row and column within that image.
Gray-level matrix image is generated by first creating the standard grayscale image from the binary, and then combining co-occurrence matrix at different rotations to form the resultant image.
Instead of standard grayscale, space-filling curves can potentially provide better basis for classification \citep{vu2020hit4mal}. Therefore, we have chosen Hilbert curves as one of the three image channels.
The entropy graph plot image used in their work captures only the block or file level entropy information whereas our entropy-based images are designed to capture file level as well as class level entropy information to identify the regions of interests with respect to both the benign and malware classes in terms of entropy.
This utilization of class level entropy information to identify such regions along with grayscale Hilbert curve is the novelty of our work.
For our few-shot implementation, we also utilize CSNN similar to \citet{conti} however that is a popular approach in literature for few-shot classification. In addition, we also perform supervised classification and HOG-PCA combination evaluation on overall detection system.

\section{Proposed Architecture for Malware Detection}\label{sec:framework}

We will now formally define the problem statement for the malware classification.

\subsection{Problem Formulation}
First, we will formulate the problem for the malware detection.
Let $\mathcal{X} = \{\mathbf{x}_{i}\}$, $i = 1,\dots,K$ , $\mathbf{x} \in \mathbb{R}^n$ be the dataset containing the input samples. This dataset contains benign samples and samples from different malware  families. The objective of the malware detection framework is to generate label $y_{i}$ for classification of data sample $\mathbf{x}_{i}$ such that:
\vspace{-2mm}
\begin{equation}
	y_{i} =
	\begin{cases}
		Benign/Malware, &\text{Binary}\\
		MalwareFamily, &\text{Multiclass}\\
	\end{cases}
\end{equation}
\noindent where $y_i$ denotes the predicted label.

\subsection{HilEnT Malware Classification Framework}

\begin{figure*}[htbp]
	\centering
	\includegraphics[width=5in]{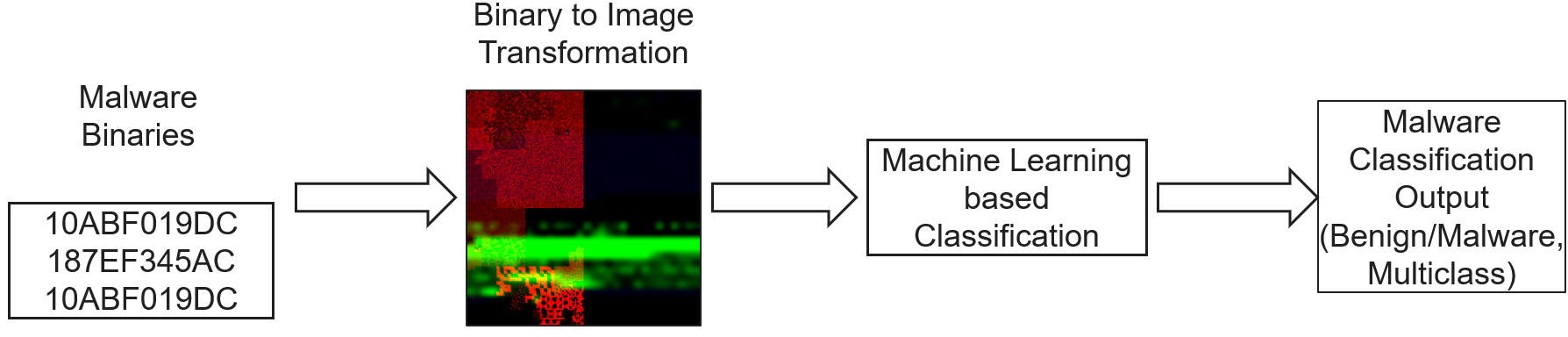}
	\caption{HilEnT Malware Classification Framework}
	\label{fig:framework}
\end{figure*}

As shown in Fig. \ref{fig:framework}, the HilEnT framework consists of two main stages: binary file to image transformation and malware classification using suitable approach. In the binary to image transformation stage, the input received was a raw binary file which contains a sequence of bytes. Hilbert curve transformation was utilized to draw a distinct pattern based on input sequence of bytes. The main motivation behind selection Hilbert curve transformation was to identify and capture byte sequences that characterize a particular malware family. By creating these locality based patterns effectively, the classifiers in the further stages of the framework pipeline can potentially better understand the features associated with malware families during training. We proposed a novel way to include entropy information within the corresponding transformed image as described in Section \ref{sec:entropy}. Finally, the combination of three grayscale or single channel images serves as a single three channel transformation or visualization of the input binary file.

In the second stage of the framework, machine learning based methods were trained to produce output as per the classification tasks.
For binary classification task, the output label will be benign or malware. For multiclass classification task, the output label will be class label for the malware family.

For the machine learning based classification, we proposed a simple CNN model and benchmarked its performance against Support Vector Machines (SVM) and Multilayer Perceptron(MLP). The simple supervised CNN classification option was designed to have minimal complexity. Combination of HOG-PCA was utilized for feature selection and dimensionality reduction to provide a time-efficient alternative with a potential accuracy trade-off in supervised classification. Additionally, we conducted few-shot learning based classification experiments to evaluate the versatility of the proposed HilEnT transformation method even in situations with limited malware samples.

From a practical application standpoint, once the machine learning based classification networks in the second stages are trained, only detection will be performed during the deployment. Hence, we have presented a detection time per sample metric during our performance evaluations to enable informed selection of detection algorithm based on application requirements.
We will now describe each stage within the framework.

\subsection{Binary to Image Transformation}
\subsubsection{Hilbert curve-based Transformation}

We opted for the Hilbert curve in our image transformation approach based on the space-filling curve technique proposed by \citet{vu2020hit4mal}. Fundamentally, this method was designed to capture subtle distinctions through improved locality retention of the malware binary data to be transformed.
Evaluating the global features of the malware binary as a whole, while reducing the pre-processing steps, was also a key motivation behind opting for this method.
Through global feature evaluation, contextual information can be assessed holistically, which may improve model robustness and reliability for practical scenarios of malware data evolving over time. 
By minimizing the pre-processing steps, computational time and resource requirements are reduced which may enable the deployment on resource-constrained clients such as edge or mobile computing devices.

Pixel color and pixel mapping are the two main components that constitute the conversion of features to images.
For a grayscale image, a value ranging from 0 to 255, is used to represent each pixel within the image; for instance, a black pixel would be represented by a value of 0, and a white pixel would be represented by a value of 255.
A grid with a preset size is then utilized to arrange these pixels.
Pixel mapping describes the arrangement of these pixels which determines the final appearance of the image.

Malware visualization method proposed in \citet{nataraj2011malware} is the most popular in literature. In this Nataraj method, PE files were first represented as sequences of sets of 8 binary values. Each set of 8 binary values was then converted to its equivalent decimal value between the range 0 to 255. This decimal value represents the grayscale pixel intensity for that set. Similarly, grayscale pixel intensity values were calculated for all the sets. For the pixel mapping step, first the width of the target image was fixed. Then, the pixels were mapped from left to right till the row width is reached, after which the pixel mapping proceeded to the next row in the image from left to right. This technique generated images with fixed width but different lengths as the number of pixels generated was determined by the input malware binary size.

According to \citet{vu2020hit4mal}, this approach may not identify the localized patterns effectively, and recommended using space-filling curves as an alternative. Space-filling curves are used for mapping one-dimensional arrays onto two-dimensional space and find common utilization in image transformation applications. Therefore, we opted for Hilbert curve as the space-filling curve with the objective of improved identification of localized patterns during transformation of PE binary files to images \citep{wijayasiri2025enhanced}.

\begin{figure}[htbp]
	\centering
	\includegraphics[width=1.2in]{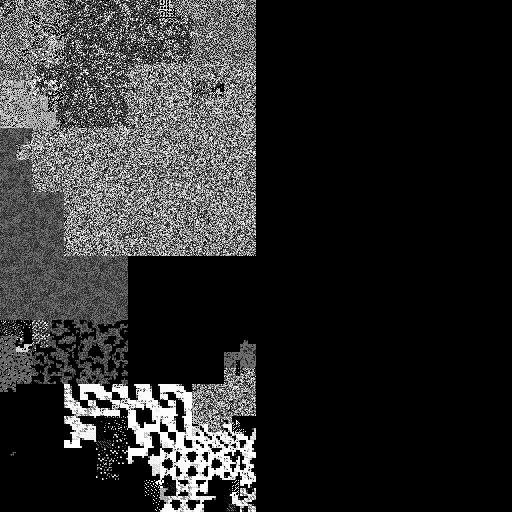}
	\caption{Hilbert curve transformed grayscale image generated from a Microsoft BIG 2015 malware sample.}
	\label{fig:sample_hilb}
\end{figure}

An example of Hilbert curve transformed image is shown in Fig. \ref{fig:sample_hilb}.

Hilbert curves are created for a block of byte sequence data within the input binary file.
As seen in the figure, by creating these locality based patterns effectively, the classifiers in the further stages of the framework pipeline can potentially better understand the features associated with malware families during training.
Length of the sequence determines the Hilbert curve pattern which is mapped to the nearest exponent of 4.
Each byte will represent a pixel and the intensity of that pixel will be given by the corresponding byte value.
Size of the generated Hilbert image is proportional to the input file size. Final image size is fixed at 256x256 and is obtained through bilinear interpolation of generated Hilbert image.
This two step process is crucial to maintain the shape of family/class specific pattern irrespective of the file size.
For example, if the file size is small, after interpolation, the pattern will be blown up along with the rest of the neighboring information. So the detection algorithm will be able to capture the overall pattern in a more meaningful way and avoid the potential issues due to varying file sizes.
The distinct patterns that potentially characterize a particular malware family may serve as key features for the classification algorithms such as CNN.
Compared to color-based Hilbert image generation in \citet{wijayasiri2025enhanced}, we adopt a grayscale Hilbert curve based image generation approach.

\subsubsection{Entropy Based Transformation}\label{sec:entropy}
Entropy is a commonly used tool in the malware detection domain. In malware binary file context, entropy can be considered as a measure of the unpredictability of the file's data. Techniques such as payload encryption or file obfuscation are used to hide the malware data. Actions such as data compression will lower the unpredictability of data, thereby raising the entropy. In general, the greater the entropy, the more likely the data is obfuscated or encrypted, and the more probable the file is malicious. Therefore, patterns of low and high entropy can be considered as regions of interest for the input sample. Our proposed entropy based novel image transformation technique is based on the observation that benign files generally have lower overall average entropy compared to malware files whereas malware regions within the malware files tend to have higher entropy than benign regions \citep{bang2024entropy}. Hence, to capture comparative information with respect to both benign and malware images within a dataset, we generated two images based on entropy values for each sample file: Benign Comparison Image and Malware Comparison Image.

For our proposed entropy transformation, we will now describe the associated key terms. Block entropy refers to the calculated entropy for a given block of byte sequence data from input sample file. Average file entropy refers to the average of entropies of all the blocks within the given binary input file. These metrics are associated with the sample entropy information. Average class entropy refers to the average of all the file entropies within the given broad class, that being the benign and malware classes. Benign Entropy cutoff $ENT_{ben}$ refers to the average class entropy of the benign class. Malware Entropy cutoff $ENT_{mal}$ refers to the average class entropy of the malware class. These two metrics are associated with the general class entropy information. Our aim was to capture and encode entropy comparison information into image transformation.

\paragraph{Sensitivity to Dataset Composition}
The entropy cut off values $ENT_{ben}$ and $ENT_{mal}$ are computed as average entropy values over the benign and malware samples in the training set, respectively. These values are therefore not fixed thresholds, but adapt naturally to the entropy characteristics of the dataset under consideration. In datasets where benign software exhibits relatively high entropy, for example due to compression or encryption, the benign cut off increases accordingly, reducing the risk of benign regions being incorrectly highlighted as malicious. Similarly, in datasets dominated by lightly obfuscated malware with lower overall entropy, the malware cut off shifts downward, preserving meaningful contrast in the malware entropy channel. In practice, we observed that relative entropy contrast, rather than the absolute cut off values themselves, plays the dominant role in discrimination. This makes HilEnT robust to moderate variations in dataset composition without requiring manual threshold tuning.

To generate these images, we first divided the file binary byte sequence into the blocks of size 256. In benign comparison image, only the pixels corresponding to blocks with entropy values below the average file entropy were populated. The intensity or pixel value of the pixel to be populated can be calculated using $((ENT_{ben}-value)/(ENT_{ben}))*255$ which captures the benign class information within the dataset via the $ENT_{ben}$ parameter.

\begin{figure}[!t]
	\centering
	\subfloat[]{\includegraphics[width=1.2in]{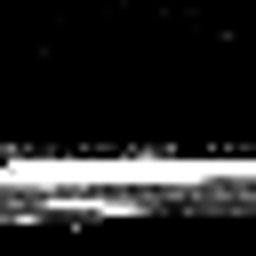}%
		\label{fig:entropy_benign}}
	\hfil
	\subfloat[]{\includegraphics[width=1.2in]{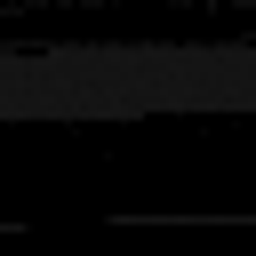}%
		\label{fig:entropy_malware}}
	\caption{Entropy cut off comparison images for a Microsoft BIG 2015 malware sample. (a) Benign entropy contrast map highlighting low entropy regions relative to the benign class average. (b) Malware entropy contrast map highlighting high entropy regions relative to the malware class average.}
	\label{fig:entropy}
\end{figure}

The pixel value represents how much lower the block entropy is compared to $ENT_{ben}$, normalized over the range $[0, ENT_{ben}]$. By extension, lower the block entropy value, larger will be the populated pixel intensity or pixel value. This process is carried out for all the blocks within the image to obtain the Benign Comparison Image. An example of Benign Comparison Image is shown in Fig.\ref{fig:entropy_benign}.

On the flip side for the Malware Comparison Image, for each block, a pixel corresponding to the block is populated only if its entropy is greater than average file entropy. The value of the populated pixel is calculated as $((value-ENT_{mal})/(8- ENT_{mal}))*255$. The intensity/pixel value is how much higher the block entropy value is compared to the $ENT_{mal}$, normalized between the range $ENT_{mal}$ to $8$. Higher the block entropy, larger will be the pixel intensity/pixel value. An example of Malware Comparison Image is shown in Fig.\ref{fig:entropy_malware} with some low intensity pixels populated in the upper half of the image. In both pixel value calculations, factor of 255 is used for scaling the grayscale image.
\subsubsection{Combined Image}
\begin{figure}[htbp]
	\centering
	\captionsetup{justification=centering}
	\includegraphics[width=1.2in]{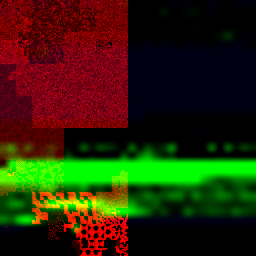}
	\caption{Final three channel HilEnT image for a Microsoft BIG 2015 malware sample obtained by combining Hilbert, benign entropy, and malware entropy channels.}
	\label{fig:sample_final}
\end{figure}

The combination of Fig.\ref{fig:sample_hilb} as red channel, Fig.\ref{fig:entropy_benign} as green channel and Fig.\ref{fig:entropy_malware} as blue channel represents the three-channel Red-Green-Blue colored (RGB) HilEnT transformed image as shown in Fig.\ref{fig:sample_final}.

\subsection{Machine Learning-based Classification}
\subsubsection{Supervised Classification: Convolutional Neural Network}
\begin{figure}[htbp]
	\centering
	\captionsetup{justification=centering}
	\includegraphics[width=2.8in]{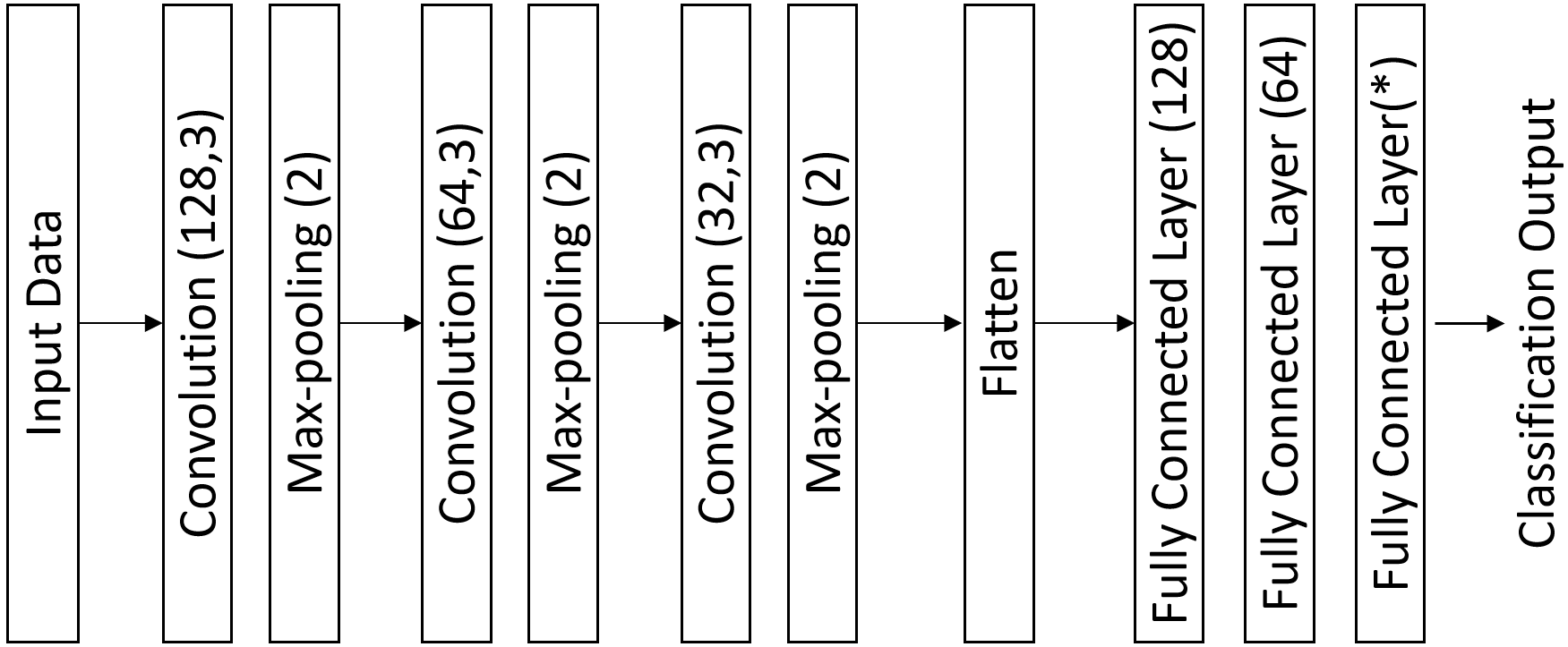}
	\caption{Lightweight CNN architecture used for supervised malware classification across all datasets.}
	\label{fig:CNNframework}
\end{figure}

For supervised classification, we proposed a CNN as shown in Fig.\ref{fig:CNNframework}
The input layer is followed by three convolutional and pooling layers.
The convolutional layer performs the process of feature extraction through convolution between data and the kernels.
Pooling layers are used to obtain higher level and more abstract features from the learned representations.
To reduce overall computations, we have utilized max-pooling in this paper to obtain the overall maximum value from the group of selected feature map values. For activation function, rectified linear unit (ReLU) will be used. The convolutional-pooling layer combinations are followed by fully-connected layers. To calculate the activation of a neuron in a fully connected layer, the weighted sum of the inputs is compared with a threshold. All activations from previous layer are connected to this neuron. The dimension of the rightmost fully connected layer denoted by (*) depends on the binary(=2) or multiclass classification(=number of classes) requirement.

\subsubsection{Histogram of Oriented Gradients (HOG)} \label{sec:hog}
The Histogram of Oriented Gradients (HOG) is a technique utilized commonly for feature extraction in image processing. HOG feature extraction process can be described as follows: First, image is divided into cells of size $S \times S$ pixels. Then orientation of each pixel is calculated. Further, these orientations are aggregated in a histogram of orientations. Finally, these histograms are combined to obtain the required feature vector. An example of HOG can be found in \citet{del2015analysis}.

\subsubsection{Principal Component Analysis (PCA)}\label{sec:pca}
Dimensionality reduction maps the data onto a space with fewer dimensions while preserving the key relationships within the data. It helps in alleviating the high dimensional data processing challenges such as more computational resources and longer processing times \citep{savakis2014efficient}. 
Additionally dimensionality reduction may help in reducing overfitting by reducing the noisy and redundant features.
Principal component analysis (PCA) reduces the dimensionality by preserving certain number of initial principal components such that majority of the variation in the original dataset is captured.
PCA is generally an effective technique for dimensionality reduction.
However, while using PCA, it is important to note that significantly reducing the dimension using PCA may potentially affect the classification performance of the machine learning based methods.
Hence, there may be a potential trade-off between processing time and classification accuracy due to dimensionality reduction.

\subsubsection{Few-shot Learning Based Classification} \label{sec:fsl}
Limited availability of samples from different malware families poses a significant challenge for malware classification \citep{wang2021novel}.
Conventional supervised machine learning based algorithms require sufficiently large amount of training data and can perform classification only within the classes or families observed during training. To enable the conventionally trained supervised model for detection of unseen or novel class, this model would need to be trained again with large number of new class samples. To address these challenges, Convolutional Siamese Neural Network (CSNN) model can be utilized.
Previous research works \citep{conti,bai2020unsuccessful,hsiao2019malware} have utilized the Siamese Neural Network to achieve good results for malware classification with limited number of samples in a dataset but with balanced distribution across different classes.

The CSNN is trained with image pairs and the pairs could either be formed by images from same family or images from different families. For a pair of images from same family, the CSNN should ideally generate the output as 1 whereas for a pair of images from different malware families, the CSNN should ideally generate the output as 0. This is based on the assumption that samples from a particular malware family generally tend to exhibit similar features or properties. During training, the CSNN is designed to identify the similarities and dissimilarities between the pairs of input data rather than each individual sample from the dataset. Structurally, CSNN can be considered as two parallel CNNs operating side by side. The weights are shared among these CNNs and the output of these two CNNs is integrated using a similarity metric. The weights are concurrently updated for both CNNs during training and Euclidean similarity is used as the similarity metric \citep{conti}.
Each CNN generates a latent representation or embedding of the corresponding image from the input image pair. Through training of CSNN, these generated latent representations become similar for image pairs from same malware family and distinct for image pairs from different malware families. As the Euclidean similarity metric compares the latent representations of these two images within a pair, higher similarity score would be obtained if the images belong to the same malware family and lower similarity score would be obtained if the images within a pair belonged to different malware families.
For few-shot learning, in general, the support set is a combined set of data which consists of reference samples obtained from each class. These reference samples serve as the representation of the corresponding classes or families. The number of instances from each class present in the support set is termed as K-shot. Query set is the set of target unlabelled samples for which the classification is to be performed. For determining the label or class of a sample from query set, a pair of the target query image and each image from the support set is provided to the CSNN. The pair which demonstrates maximum similarity score, implies that the query image is similar to support set image within that pair. Therefore, the family class of that support image is identified as the family class of the query image.
As CSNN does not require significant amount of training data\citep{conti}, it addresses the limited data availability issue. Additionally, if at least one sample of a novel class is available in the support set, CSNN can identify similar samples belonging to that class within the query set based on similarity score without needing to train using that class.

\section{Experimental Setup}\label{sec:setup}
\subsection{Datasets}
The four datasets used for performance evaluation will now be described in brief.
\subsubsection{Dike Dataset}
Dike Dataset is a labelled dataset which contains benign and malicious Portable Executable (PE) files and Object Linking and Embedding (OLE) files\citep{dikedataset}. There are 1082 benign files (982 PE, 100 OLE) and 10841 malicious files (8970 PE, 1871 OLE). The samples for this dataset are obtained from sources such as MalwareBazaar, DuckDuckGo. In this work, we are only considering the PE files from both benign and malicious types. Additionally, we are using this dataset only for binary classification experiment and hence only benign and malicious labels are considered.

\subsubsection{Michael Lester PE Dataset}
This dataset was created by sampling PE files from PE Malware Dataset by Michael Lester which was published by Practical Security Analytics\citep{lesterpratical}.
The dataset is primarily sourced from VirusShare, MalShare and TheZoo.
The dataset contains around 200000 samples with 86812 benign and 114737 malicious samples. We randomly sampled 10000 malicious and 10000 benign files from the total dataset to create a balanced dataset with comparable size to rest of our datasets. This will also be used only for binary classification experiment and as such only benign and malicious labels are considered.

\subsubsection{Microsoft BIG 2015}
Microsoft introduced this dataset in 2015 in a Kaggle competition\citep{ronen2018microsoft}.
There are total 21,741 malware samples in the dataset which are further categorized as a labelled training set of 10,868 samples and an unlabelled set of 10,873 samples. In this work, we have only utilized the labelled training set, which consists of 9 malware families. Additionally, we have only utilized the .bytes files provided in the dataset. Furthermore, some files within the dataset contain an unrecognizable symbol represented by the ``??" in the byte sequence. Some files containing only the unrecognizable symbol were removed from the dataset. The different classwise split for the dataset is shown in the Table \ref{table:msbig_samp_distrib}.

\begin{table}[htbp]
	\centering
	\caption{Sample Distribution for MicrosoftBIG Dataset}
	\label{table:msbig_samp_distrib}
	{\begin{tabular}{|c|>{\centering}m{1cm}|c|}
			\hline
			\textbf{Family} & \textbf{Sample Count} & \textbf{Distribution(\%)} \\
			\hline
			Ramnit & 1541 & 14.2 \\
			\hline
			Lollipop & 2478 & 22.8 \\
			\hline
			Kelihos\_v3 & 2942 & 27.1 \\
			\hline
			Vundo & 475 & 4.4 \\
			\hline
			Simda & 42 & 0.4 \\
			\hline
			Tracur & 751 & 6.9 \\
			\hline
			Kelihos\_v1 & 398 & 3.6 \\
			\hline
			Obfuscator & 1228 & 11.3 \\
			\hline
			Gatak & 1013 & 9.3 \\
			\hline
			\hline
			\textbf{Total} & 10868 & 100 \\
			\hline			
	\end{tabular}}
\end{table}

\subsubsection{Self-collected Dataset}
PE files are the main focus of our experiment. During the PE file collection for self-collected dataset, the size limitation is set to 5 MB for both malware and benign samples. In line with the static analysis procedure described in \citet{krvcal2018deep}, packed malware samples are excluded from the dataset by using Detect It Easy\citep{detectiteasy}. The dataset contains 4549 malware samples sourced from the well-known malware database MalwareBazaar, spanning the years 2020 to 2023. 
VirusTotal, a widely-used commercial malware detection service that utilizes 72 malware detectors, is further employed to scan all the malware samples, retaining those detected by more than 40 engines. The malware samples encompass total 8 malware families. The sample distribution for the self collected dataset is shown in Table \ref{table:self_samp_distrib}.

\begin{table}[htbp]
	\centering
	\caption{Sample Distribution for Self Collected Dataset}
	\label{table:self_samp_distrib}
	{\begin{tabular}{|c|>{\centering}m{1cm}|c|}
			\hline
			\textbf{Family} & \textbf{Sample Count} & \textbf{Distribution(\%)} \\
			\hline
			GuLoader & 2053 & 45.2 \\
			\hline
			RedLineStealer & 493 & 10.8 \\
			\hline
			Heodo & 464 & 10.2 \\
			\hline
			TrickBot & 351 & 7.7 \\
			\hline
			Loki & 341 & 7.5 \\
			\hline
			GrandCrab & 340 & 7.5 \\
			\hline
			SmokeLoader & 265 & 5.8 \\
			\hline
			IcedID & 242 & 5.3 \\
			\hline
			\hline
			\textbf{Total} & 4549 & 100 \\
			\hline			
	\end{tabular}}
\end{table}

From the datasets under evaluation, Microsoft Dataset and Self-collected Dataset did not have their own benign samples for binary classification. Therefore, we utilized the same set of benign samples from the Lester PE dataset as those samples are publicly available. Also, the count of selected benign samples from the Lester PE dataset is roughly similar to Microsoft Dataset which provides a balanced dataset for binary classification.

\subsection{Evaluation Metrics}
For the evaluation of malware detection, we have adopted accuracy metric for binary and multiclass classification performance comparison as we have a balanced dataset with similar number of normal and malicious samples. For a fair comparison, the training set and test set will be kept identical for each method under the evaluation.
Experiment-specific dataset splits are explained in the corresponding relevant experiments in Section \ref{sec:results}. In certain experiments, we also provided detection time per sample metric to evaluate the impact of HOG-PCA step on overall time performance. 

\subsection{System Setup}
Sklearn\citep{scikit-learn}, Keras\citep{chollet2015keras} and PyTorch\citep{paszke2019pytorch} were utilized for the implementation of the proposed framework in Python. The evaluation system specifications include: Intel i7-10700K CPU, 64GB RAM, Nvidia RTX3070 GPU(8GB) and Ubuntu 20.04 OS.

\paragraph{Implementation Details for Reproducibility}
For all supervised experiments, we used the same lightweight CNN architecture shown in Fig.~\ref{fig:CNNframework}. The network consists of three convolutional layers with $3 \times 3$ kernels and ReLU activations, each followed by max pooling, and three fully connected layers for classification. Training was performed using the Adam optimizer with a learning rate of $1 \times 10^{-3}$ and a batch size of 32. For feature based experiments, Histogram of Oriented Gradients was computed using $8 \times 8$ pixel cells with block normalization, and Principal Component Analysis was applied to retain 95\% of the cumulative variance. All parameters were kept fixed across datasets and experiments to ensure fair comparison and reproducibility.

\section{Experiments and Results}\label{sec:results}
The performance of our proposed HilEnT framework was evaluated across three main experiments: Binary Classification, Multiclass Classification and Unseen Class Classification experiment. It is important to note that binary classification here refers to classification with only two classes (benign and malware).
Supervised classification setting was used to produce either some or all the results within all three experiments.
On the other hand, few-shot setting was used to perform only the multiclass classification and unseen class classification experiment. Within the supervised classification setting, we generated two result points for each possible case: with and without HOG-PCA. Within the few-shot setting, we have evaluated two support set sizes: 1 and 10. To provide a baseline comparison for our proposed HilEnT method, we generated Nataraj malware visualization image\citep{nataraj2011malware} for certain experiments. In addition to proposed CNN, we also implemented two popular machine learning algorithms: Multilayer Perceptron (MLP) and Support Vector Machines (SVM) for baselines. MLP is constructed using multiple layers of basic processing units called perceptrons. We will now describe our individual experiments and discuss their results.

\subsection{Binary Classification}\label{sec:BinaryClass}
For this experiment, samples from different malware types were considered under single malicious category. As Dike and Lester PE (Lester) are binary datasets, no further processing was required. For Microsoft BIG (MS) and Self Collected (Self) Dataset, samples from different malware families were considered as a single malicious category. Only simple supervised and HOG-PCA enhanced supervised learning based experiment is carried out for binary classification experiment. The few-shot learning based method was not used for the binary classification experiment. For this experiment, 5-fold cross validation was used with final results obtained using average of 5-fold results.
We also discuss the detection time per sample for all methods under consideration in this binary classification task.  

The results are summarized in Table \ref{table:exp1_binary}. The 'Image Txm Method' refers to the Image Transformation method used to convert the byte sequence into image files. `NO' HOG + PCA rows in the table refer to the simple supervised learning case. The SVM (Gray), MLP (Gray) and CNN (Gray) are used for establishing baseline for the Hilbert curve or Nataraj transformed grayscale images and therefore, these methods are not utilized for the HilEnT transformed images. The results for the HilEnT method are shown in the separate column for easier comparison with the remaining models.

\begin{table*}[htbp]
	\centering
	\caption{Experiment 1: Binary classification accuracy for Hilbert and Nataraj baselines with and without HOG and PCA, compared against the proposed HilEnT method. Results are averaged over five fold cross validation using balanced benign and malware splits.}
	\label{table:exp1_binary}
	{\begin{tabular}{|>{\centering\arraybackslash}m{1.3cm}|>{\centering\arraybackslash}m{1.3cm}|>{\centering\arraybackslash}m{1cm}|>{\centering\arraybackslash}m{1cm}|>{\centering\arraybackslash}m{1cm}|>{\centering\arraybackslash}m{1cm}|>{\centering\arraybackslash}m{1.3cm}|}%
					\cline{1-7}
					& \textbf{Image} & \textbf{HOG} & \multicolumn{4}{|c|}{\textbf{Accuracy}} \\ 
					\cline{4-7}
					\textbf{Dataset} & \textbf{Txm} & \textbf{+} & \textbf{SVM} & \textbf{MLP} & \textbf{CNN} & \multirow{2}{*}{\textbf{HilEnT}} \\%
					& \textbf{Method} & \textbf{PCA} & (Gray) & (Gray) & (Gray) &  \\%
					\hline
					\multirow{4}{*}{Dike} & \multirow{2}{*}{Hilbert} & YES & 0.97 & 0.98 & 0.97 & \multirow{4}{*}{\textbf{0.99}}\\%
					\cline{3-6} %
					& & NO & 0.98 & 0.98 & 0.98 & \\%
					\cline{2-6} %
					& \multirow{2}{*}{Nataraj} & YES & 0.97 & 0.98 & 0.97 & \\%
					\cline{3-6} %
					& & NO & 0.98 & 0.98 & 0.98 & \\%
					\hline

					\multirow{4}{*}{Lester} & \multirow{2}{*}{Hilbert} & YES & 0.85 & 0.85 & 0.84 & \multirow{4}{*}{\textbf{0.94}} \\%
					\cline{3-6} %
					& & NO & 0.86 & 0.84 & 0.88 & \\%
					\cline{2-6} %
					& \multirow{2}{*}{Nataraj} & YES & 0.85 & 0.84 & 0.83 & \\%
					\cline{3-6} %
					& & NO & 0.86 & 0.85 & 0.88 & \\%
					\hline

					\multirow{4}{*}{MS} & \multirow{2}{*}{Hilbert} & YES & 0.93 & 0.93 & 0.92 & \multirow{4}{*}{\textbf{0.99}} \\%
					\cline{3-6} %
					& & NO & 0.95 & 0.96 & 0.97 & \\%
					\cline{2-6} %
					& \multirow{2}{*}{Nataraj} & YES & 0.93 & 0.93 & 0.91 & \\%
					\cline{3-6} %
					& & NO & 0.93 & 0.94 & 0.96 & \\%
					\hline

					\multirow{4}{*}{Self} & \multirow{2}{*}{Hilbert} & YES & 0.95 & 0.96 & 0.94 & \multirow{4}{*}{\textbf{0.99}} \\%
					\cline{3-6} %
					& & NO & 0.94 & 0.96 & 0.97 & \\%
					\cline{2-6} %
					& \multirow{2}{*}{Nataraj} & YES & 0.95 & 0.96 & 0.92 & \\%
					\cline{3-6} %
					& & NO & 0.94 & 0.96 & 0.97 & \\%
					\hline						
			\end{tabular}}
		\end{table*}

\begin{table*}
	\small
	\centering
	\caption{Experiment 1: Per sample detection time for binary classification using Hilbert and Nataraj baselines with and without HOG and PCA, compared against the proposed HilEnT method.}
	\label{table_appendix:exp1_binary}
			{\begin{tabular}{|>{\centering\arraybackslash}m{1.3cm}|>{\centering\arraybackslash}m{1.3cm}|>{\centering\arraybackslash}m{1cm}|>{\centering\arraybackslash}m{1cm}|>{\centering\arraybackslash}m{1cm}|>{\centering\arraybackslash}m{1cm}|>{\centering\arraybackslash}m{1.3cm}|}%
							\cline{1-7}
							& \textbf{Image} & \textbf{HOG} & \multicolumn{4}{|c|}{\textbf{Detection Time per sample(sec)}} \\
							\cline{4-7}
							\textbf{Dataset} & \textbf{Txm} & \textbf{+} & \textbf{SVM} & \textbf{MLP} & \textbf{CNN} & \multirow{2}{*}{\textbf{HilEnT}} \\%
							& \textbf{Method} & \textbf{PCA} & (Gray) & (Gray) & (Gray) &  \\%
							\hline
							\multirow{4}{*}{Dike} & \multirow{2}{*}{Hilbert} & YES & 4.1e-5 & 6e-7 & 2.8e-5 & \multirow{4}{*}{6.1e-4} \\
							\cline{3-6} 
							& & NO & 0.34 & 5.1e-5 & 9.1e-3 & \\
							\cline{2-6} 
							& \multirow{2}{*}{Nataraj} & YES & 3.7e-5 & 5.8e-7 & 2.5e-5 & \\
							\cline{3-6}
							& & NO & 0.26 & 4.3e-5 & 9.2e-3 & \\
							\hline

							\multirow{4}{*}{Lester} & \multirow{2}{*}{Hilbert} & YES & 6.6e-4 & 2.7e-7 & 3.3e-4 & \multirow{4}{*}{4.7e-4} \\
							\cline{3-6}
							& & NO & 0.59 & 3.3e-5 & 9.5e-3 & \\
							\cline{2-6}
							& \multirow{2}{*}{Nataraj} & YES & 4.1e-4 & 2.7e-7 & 3.2e-4 & \\
							\cline{3-6}
							& & NO & 0.39 & 3.2e-5 & 9.6e-3 & \\
							\hline

							\multirow{4}{*}{MS} & \multirow{2}{*}{Hilbert} & YES & 2.5e-4 & 2.3e-7 & 3e-5 & \multirow{4}{*}{4.2e-4} \\
							\cline{3-6}
							& & NO & 0.42 & 5.6e-5 & 8.9e-3 & \\
							\cline{2-6}
							& \multirow{2}{*}{Nataraj} & YES & 2.6e-4 & 2.7e-7 & 3.2e-5 & \\
							\cline{3-6}
							& & NO & 0.29 & 7.2e-5 & 9e-3 & \\
							\hline

							\multirow{4}{*}{Self} & \multirow{2}{*}{Hilbert} & YES & 1.2e-4 & 2.4e-7 & 3.2e-4 & \multirow{4}{*}{5.4e-4} \\
							\cline{3-6}
							& & NO & 0.47 & 6.1e-5 & 2.1e-3 & \\
							\cline{2-6}
							& \multirow{2}{*}{Nataraj} & YES & 1.4e-4 & 2.3e-7 & 3.5e-4 & \\
							\cline{3-6}
							& & NO & 0.35 & 5.1e-5 & 2.2e-3 & \\
							\hline									
							
					\end{tabular}}
				\end{table*}

As seen from the Table \ref{table:exp1_binary}, the HilEnT displays best performance across different datasets for binary classification. Generally, combination of HOG and PCA tends to show similar accuracy performance irrespective of the transformation method. However, we observed that the HOG-PCA combination is around two orders of magnitude faster in detection time per sample metric across all datasets compared to the methods not using PCA as seen in Table \ref{table_appendix:exp1_binary}. Therefore, as intended, a trade-off can be achieved between accuracy and detection time performance with pre-processing step of HOG-PCA combination. SVM performance was mostly among the slowest across different datasets and methods.
MLP and CNN performance is generally on par with each other in most cases.
Nataraj and Hilbert curve transformation methods mostly perform on par as well for most cases both in terms of accuracy and detection time performance.
Among the four datasets, lowest binary classification accuracy performance was observed for Lester dataset across all machine learning and image transformation methods.

\subsection{Multiclass Classification}

For this experiment, samples from different malware families were considered under their own malicious family category. As Dike and Lester PE (Lester) are binary datasets, those datasets are not considered for the multiclass experiment. All three algorithms: simple supervised and HOG-PCA enhanced supervised learning and few-shot learning based experiment is carried out for multiclass classification experiment. For this experiment, the 80\% of main dataset was used for training and the remaining 20\% of main dataset was used for test data. We did not report the detection time per sample for this experiment as it is similar to the results obtained in Experiment 1. Additionally, since SVM does not have an inherent multiclass classification option and SVM time performance in previous experiment was significantly slower compared to the other methods, we did not report the results for the SVM in multiclass classification experiment.

The results are summarized in Table \ref{table:exp2_multi}. The 'Image Txm Method' refers to the Image Transformation method used to convert the byte sequence into image files. `NO' HOG + PCA rows in the table refer to the simple supervised learning case. The results for the HilEnT method and HilEnT with Few-shot learning method are shown in the separate columns for easier comparison with the remaining models. The MLP (Gray) and CNN (Gray) indicate that the image transformation method for those algorithms consisted only of Hilbert curve or Nataraj transformed image exclusively without the entropy based transformed images.

\begin{table*}[htbp]
	\centering
	\caption{Experiment 2: Multiclass classification accuracy for Hilbert and Nataraj baselines with and without HOG and PCA, compared against the proposed HilEnT and few shot learning methods.}
	\label{table:exp2_multi}
			{\begin{tabular}{|>{\centering\arraybackslash}m{1.3cm}|>{\centering\arraybackslash}m{1.3cm}|>{\centering\arraybackslash}m{1cm}|>{\centering\arraybackslash}m{1.28cm}|>{\centering\arraybackslash}m{1.2cm}|>{\centering\arraybackslash}m{1.2cm}|>{\centering\arraybackslash}m{1.2cm}|}%
							\cline{1-7}
							& \textbf{Image} & \textbf{HOG} & \multicolumn{4}{|c|}{\textbf{Accuracy}} \\
							\cline{4-7}
							\textbf{Dataset} & \textbf{Txm} & \textbf{+} & \textbf{MLP} & \textbf{CNN} & \multirow{2}{*}{\textbf{HilEnT}} & \textbf{Few-} \\
							& \textbf{Method} & \textbf{PCA} & (Gray) & (Gray) & & \textbf{Shot}\\							
							\hline
							\multirow{4}{*}{MS} & \multirow{2}{*}{Hilbert} & YES & 0.93 & 0.93 & \multirow{4}{*}{\textbf{0.97}} & \multirow{4}{*}{0.95} \\
							\cline{3-5}
							& & NO & 0.90 & 0.91 & & \\
							\cline{2-5}
							& \multirow{2}{*}{Nataraj} & YES & 0.94 & 0.94 & & \\
							\cline{3-5}
							& & NO & 0.91 & 0.93 & & \\
							\hline

							\multirow{4}{*}{Self} & \multirow{2}{*}{Hilbert} & YES & 0.91 & 0.92 & \multirow{4}{*}{\textbf{0.95}} & \multirow{4}{*}{0.92} \\
							\cline{3-5}
							& & NO & 0.90 & 0.91 & & \\
							\cline{2-5}
							& \multirow{2}{*}{Nataraj} & YES  & 0.92 & 0.93 & & \\
							\cline{3-5}
							& & NO & 0.90 & 0.92 & & \\
							\hline							 
							
					\end{tabular}}
				\end{table*}
			
As seen from the Table \ref{table:exp2_multi}, the HilEnT displays best performance across both datasets for multiclass classification. Unlike binary classification, combination HOG and PCA generally demonstrates higher accuracy performance irrespective of the transformation method for grayscale image based multiclass classification. This could be attributed to the relevant feature extraction via HOG for multiclass classification.
Nataraj and Hilbert curve transformation methods mostly perform on par for most cases both in terms of accuracy performance. Few-shot learning based classification shows only marginally lower accuracy performance despite being trained with only fraction of data (around 10\%) compared to other supervised methods. It indicates that few-shot based learning algorithm was able to understand the relevant class comparison patterns from limited samples available. However, the performance difference is slightly larger for self collected dataset.
		
\subsection{Unseen Class Classification}\label{sec:unseen}
Along with existing known families or types of malware, many new malware emerge regularly, and it is not practical to obtain training data for all such malware. Hence unseen class detection is an important set of evaluations to identify how the malware detection models perform when unseen or novel class samples e.g. zero-day attacks are presented for detection. This experiment is performed only with simple supervised learning and few-shot learning (FSL) based methods.
Though we are displaying the results for both methods in the same table, there is a small difference between the method in which this experiment is performed for simple supervised learning and FSL. As Dike and Lester PE are binary datasets, those datasets are not considered for the unseen class experiment. To perform this set of experiments for simple supervised learning, during each round of the experiment, one of the malware family classes was removed from main dataset to serve exclusively as the test set. The remaining dataset, containing benign samples and malicious samples from the remaining classes, formed the training set. The training set was categorized in binary classes i.e. benign and malicious samples. On the other hand, the entire test set contained only malicious samples, as all the samples belong to the single unseen malware family class.
		
To perform this set of experiments for the few-shot learning based setting, during each round of experiment, one of the malware classes was treated as unseen class. Only 150 samples from each seen class are available for training. No samples from unseen class are used during training. During testing, support set consisted of samples from the all the classes. Only the unseen class samples were present in the query set during testing. Hence, unlike simple supervised setting, benign samples are not used. 
Two support set sizes were considered: 1 (Sup=1) and 10 (Sup=10). The results reported here are accuracy/recall value for the unseen class.

				\begin{table*}
					\centering
					\caption{Experiment 3: Unseen malware class detection accuracy for the Microsoft BIG 2015 and self collected datasets using supervised CNN, few shot learning with one and ten support samples, and the proposed HilEnT method.}
					\label{table:exp3_unseen}
									{\begin{tabular}{|c|c|c|c|c|}%
											\cline{1-5}
											\multicolumn{5}{|c|}{\textbf{a. Accuracy for Microsoft BIG Dataset}} \\
											\hline
											\multirow{2}{*}{\textbf{Unseen Class}} & \textbf{CNN} & \multirow{2}{*}{\textbf{HilEnT}} & \textbf{FSL} & \textbf{FSL} \\ %
											& (Hilbert) & &(Sup=1) & (Sup=10) \\
											\cline{1-5}
											Ramnit & 0.96 & 0.72 & 0.62 & 0.57 \\
											Lollipop & 0.76 & 0.82 & 0.68 & 0.39  \\ 
											Kelihos\_v3 & 0.85 & 0.99 & 1.00 & 1.00  \\ 
											Vundo & 0.99 & 0.99 & 0.66 & 0.63 \\ 
											Simda & 0.98 & 0.97 & 0.66 & 0.69 \\ 
											Tracur & 0.99 & 0.97 & 0.39 & 0.30 \\ 
											Kelihos\_v1 & 0.99 & 0.93 & 0.88 & 0.93 \\ 
											Obfuscator & 0.99 & 0.94 & 0.77 & 0.71 \\
											Gatak & 0.99 & 0.98 & 0.73 & 0.64 \\
											\hline
											Weighted Acc & 0.89 & 0.90 & 0.77 & 0.67 \\
											\hline
											\hline					
											\multicolumn{5}{|c|}{\textbf{b. Accuracy for self-collected Dataset}} \\
											\hline
											\multirow{2}{*}{\textbf{Unseen Class}} & \textbf{CNN} & \multirow{2}{*}{\textbf{HilEnT}} & \textbf{FSL} & \textbf{FSL} \\ %
											& (Hilbert) & &(Sup=1) & (Sup=10) \\
											\cline{1-5}
											GuLoader & 0.94 & 0.86 & 0.62 & 0.84 \\
											RedLineStealer & 0.82 & 0.86 & 0.86 & 0.83  \\ 
											Heodo & 0.99 & 0.93 & 0.81 & 0.81  \\ 
											TrickBot & 0.28 & 0.96 & 0.87 & 0.88 \\ 
											Loki & 0.93 & 0.96 & 0.89 & 0.89 \\ 
											GrandCrab & 0.75 & 0.99 & 0.89 & 0.90 \\ 
											SmokeLoader & 0.97 & 0.98 & 0.86 & 0.90 \\ 
											IcedID & 0.53 & 0.25 & 0.90 & 0.8 \\					
											\hline
											Weighted Acc & 0.88 & 0.87 & 0.72 & 0.85 \\
											\hline
									\end{tabular}}
								\end{table*}
								
The results for this experiment are shown in Table \ref{table:exp3_unseen}. For simple supervised setting, performance for both grayscale Hilbert curve and HilEnT image is consistent across different classes. Similar observations can be made for self-collected dataset performance. However, for IcedID family, HilEnT performs significantly lower compared to other methods.

For few-shot experiment, the results for the MicrosoftBIG dataset show inconsistent results across different support set sizes. Kelihos\_v3 and Kelihos\_v1 are overall easier to identify. Tracur family samples seems to be most difficult to classify when they are unseen. For self collected, overall consistent results across different support set sizes are observed except for class 1 with support set size 1.
								
Overall drop in performance for unseen class experiment is anticipated. Simple supervised method has only two possible classes as the output for unseen data: benign or malware. Therefore, potential for missclassifcation is small. On the other hand, in few-shot learning setting, the number of possible output classes depends on the number of classes present in the support set which can reach up to total number of known classes within the dataset. Thus the potential to misclassify is higher in few-shot case owing to its inherent setup.

This unseen class experiment enabled us to test our proposed methodologies for robustness against unseen class data.

\subsection{State-of-the-art (SoTA) Comparison}
Table \ref{table:sota} demonstrates the performance comparison of our proposed model with some SoTA works for Microsoft BIG 2015 dataset. For each comparison, we have replicated the train-test split or number of benign samples utilized. As these SoTA models are not publicly available, we have directly compared our results with the results from their corresponding paper. Cells with `NA' entries in Table \ref{table:sota} indicate that binary classification results were not reported for the corresponding SoTA work. Also note that, as Microsoft BIG 2015 dataset does not provide any benign samples, each SoTA work uses their own set of benign files.

\begin{table*}[htbp]
	\centering
	\caption{Accuracy comparison between the proposed HilEnT method and prior state of the art approaches on the Microsoft BIG 2015 malware dataset.}
	\label{table:sota}
	{\begin{tabular}{|>{\centering\arraybackslash}m{4cm}|>{\centering\arraybackslash}m{2.5cm}|>{\centering\arraybackslash}m{1cm}|>{\centering\arraybackslash}m{1.28cm}|}%
			\hline
			& & \multicolumn{2}{|c|}{\textbf{Accuracy}} \\ %
			\cline{3-4}
			\textbf{SoTA} & \textbf{Classification}  & \textbf{SoTA} & \textbf{HilEnT} \\
			\hline
			\multirow{2}{*}{\citep{lo2019xception}} & Binary & NA & NA \\
			\cline{2-4}
			& Multiclass & 0.992 & 0.985 \\
			\hline
			\multirow{2}{*}{\citep{hemalatha2021efficient}} & Binary & 0.977 & 0.991 \\
			\cline{2-4}
			& Multiclass & 0.985 & 0.973 \\
			\hline
			\multirow{2}{*}{\citep{conti}} & Binary & NA & NA \\
			\cline{2-4}
			& Multiclass & 0.986 & 0.979 \\
			\hline
	\end{tabular}}
\end{table*}
								
Our proposed HilEnT achieves comparable results to the SoTA. It performs better in binary classification but performs slightly lower in multiclass classification tasks.

\citet{hemalatha2021efficient} uses DenseNet network which consists of at least 98 convolution layers based on the network information provided in their paper. Xception model used by \citet{lo2019xception} uses at least 36 convolutional layers. Our proposed CNN network uses 3 convolutional layers followed by 3 fully connected layers. Thus, the number of total layers in our network can be considered 6.

Focus for \citet{hemalatha2021efficient} and \citet{lo2019xception} was using deeper convolutional networks for malware classification. Those works have used standard approach to convert malware binaries to images. Our focus was to propose a novel method to transform malware binary to images such that even shallow CNNs are able to perform the classification. Our proposed method utilizes combination of Hilbert curve transformation and entropy based transformations to extract and represent the key features directly during transformation. This step enables our approach to utilize shallower networks for malware classification. Despite using shallower convolutional neural networks, we can achieve comparable performance for multiclass classification and even better performance for binary classification than the SoTA. Using shallower networks generally lowers the computational requirements and improves the detection time performance.

Compared to algorithm proposed by \citet{conti}, we proposed a different file binary to image transformation technique but still achieved comparable results. Our proposed method encodes benign and malware class entropy information within the transformation process which is novel to the best of our knowledge. This novel approach which includes class entropy comparison has a potential to provide fairer entropy based comparison in cases where the malware families within certain datasets exhibit atypical behaviour. For instance, unlike general assumption where benign files exhibit lower entropy values, if a dataset contains benign files with high entropy values, this class behaviour will be captured under benign comparison images through class threshold for our proposed HilEnT method.

\begin{table*}[htbp]
\centering
\caption{Backbone Comparison of Malware Visualization Methods: Nataraj (baseline) vs. HilEnT (ours). Higher accuracy is bolded; improvements greater than 1\% absolute accuracy are marked with *.}
\label{table:backbone_comparison}
\small
\begin{tabular}{|c|c|c|c|}
\hline
\textbf{Dataset} & \textbf{Backbone Model} & \textbf{Nataraj Acc.} & \textbf{HilEnT Acc.} \\
\hline

\multirow{10}{*}{\textbf{Dike}} 
& AlexNet              & 0.99297 & \textbf{0.99448} \\
& VGG16                & 0.98995 & \textbf{0.99648}\textsuperscript{*} \\
& VGG19                & 0.98995 & \textbf{0.99498} \\
& ResNet50             & 0.99146 & \textbf{0.99397} \\
& ResNet152            & 0.99196 & \textbf{0.99397} \\
& DenseNet121          & \textbf{0.99498} & 0.99448 \\
& EfficientNet-B0      & 0.99046 & \textbf{0.99548}\textsuperscript{*} \\
& MobileNetV3-Large    & 0.99347 & \textbf{0.99598} \\
& ConvNeXt-Tiny        & 0.99347 & \textbf{0.99598} \\
& ViT-B/16             & 0.98744 & \textbf{0.99046} \\
\hline

\multirow{10}{*}{\textbf{Lester}} 
& AlexNet              & 0.94318 & \textbf{0.94988} \\
& VGG16                & 0.95533 & \textbf{0.95906} \\
& VGG19                & 0.93722 & \textbf{0.95757}\textsuperscript{*} \\
& ResNet50             & 0.94541 & \textbf{0.96005}\textsuperscript{*} \\
& ResNet152            & 0.94591 & \textbf{0.95906} \\
& DenseNet121          & 0.95484 & \textbf{0.96154} \\
& EfficientNet-B0      & 0.95211 & \textbf{0.96129} \\
& MobileNetV3-Large    & 0.93995 & \textbf{0.95931}\textsuperscript{*} \\
& ConvNeXt-Tiny        & 0.95583 & \textbf{0.96501} \\
& ViT-B/16             & 0.92854 & \textbf{0.93871} \\
\hline

\multirow{10}{*}{\textbf{MS (Microsoft)}} 
& AlexNet              & 0.98802 & \textbf{0.99449} \\
& VGG16                & 0.99521 & \textbf{0.99808} \\
& VGG19                & \textbf{0.99784} & \textbf{0.99784} \\ 
& ResNet50             & 0.99664 & \textbf{0.99880} \\
& ResNet152            & 0.99688 & \textbf{0.99856} \\
& DenseNet121          & 0.99616 & \textbf{0.99880} \\
& EfficientNet-B0      & 0.99688 & \textbf{0.99880} \\
& MobileNetV3-Large    & 0.99425 & \textbf{0.99832}\textsuperscript{*} \\
& ConvNeXt-Tiny        & 0.99784 & \textbf{0.99832} \\
& ViT-B/16             & 0.98873 & \textbf{0.99616}\textsuperscript{*} \\
\hline

\multirow{10}{*}{\textbf{Self-Collected}} 
& AlexNet              & 0.98591 & \textbf{0.99107} \\
& VGG16                & 0.98591 & \textbf{0.99416}\textsuperscript{*} \\
& VGG19                & 0.98832 & \textbf{0.99381} \\
& ResNet50             & 0.99278 & \textbf{0.99485} \\
& ResNet152            & 0.99141 & \textbf{0.99549} \\
& DenseNet121          & 0.99072 & \textbf{0.99513} \\
& EfficientNet-B0      & 0.99210 & \textbf{0.99401} \\
& MobileNetV3-Large    & 0.99141 & \textbf{0.99462} \\
& ConvNeXt-Tiny        & \textbf{0.99381} & 0.99395 \\ 
& ViT-B/16             & 0.97629 & \textbf{0.99086}\textsuperscript{*} \\
\hline

\end{tabular}
\end{table*}

As shown in Table~\ref{table:backbone_comparison}, the performance of HilEnT is generally comparable to the standard Nataraj grayscale visualization across all evaluated CNN backbones and datasets. Bold values indicate the higher accuracy within each pairwise comparison, and in most cases the improvements are incremental and consistent with normal architectural variability. However, entries marked with a superscript * denote cases where the absolute improvement exceeds 1\%, indicating a more notable increase in discriminative capability for certain model–dataset combinations. These moderate but meaningful gains demonstrate that while both visualization methods can support effective CNN-based malware detection, HilEnT introduces clearer structural cues that particularly benefit some backbones without disrupting overall performance stability.

\begin{table}[htbp]
\centering
\caption{Per-sample inference time (seconds) on the self-collected dataset across several CNN backbones using the Nataraj visualization, compared with our HilEnT method.}
\label{table:runtime_comparison}
\small
\begin{tabular}{|c|c|c|}
\hline
\textbf{Dataset} & \textbf{Model} & \textbf{Time (s)} \\
\hline
\multirow{10}{*}{\textbf{Self}} 
& AlexNet              & 8.81e-4 \\
& VGG16                & 2.01e-3 \\
& VGG19                & 2.73e-3 \\
& ResNet50             & 1.52e-3 \\
& ResNet152            & 3.15e-3 \\
& DenseNet121          & 1.82e-3 \\
& EfficientNet-B0      & 1.06e-3 \\
& MobileNetV3-Large    & 9.86e-4 \\
& ConvNeXt-Tiny        & 2.03e-3 \\
& ViT-B/16             & 4.83e-3 \\
\hline
\multicolumn{2}{|c|}{\textbf{HilEnT (Ours)}} & \textbf{5.4e-4} \\
\hline
\end{tabular}
\end{table}

Although HilEnT achieves slightly lower accuracy on the Lester dataset compared to some of the deeper CNN backbones evaluated in Table~\ref{table:backbone_comparison}, this reduction is modest and occurs primarily in cases where the competing models rely on substantially larger architectures. In contrast, Table~\ref{table:runtime_comparison} demonstrates that HilEnT offers a clear advantage in inference efficiency, processing samples in 5.4e-4 seconds, significantly faster than any of the CNN-based models. The runtimes across the compared backbones show that deeper networks such as VGG19, ResNet152, and ViT-B/16 incur noticeably higher computational cost, while even the lightweight models remain slower than HilEnT. This highlights a favorable trade off: HilEnT delivers competitive accuracy across all datasets while achieving the lowest inference time, making it particularly suitable for real-time or resource-constrained malware detection scenarios.

\begin{table*}[htbp]
\centering
\caption{Performance comparison of CNN-AutoMIC using Nataraj visualization versus our HilEnT visualization across four datasets. Bold values indicate the higher score for each metric.}
\label{table:cnnautomic_comparison}
\small
\begin{tabular}{|c|c|ccccc|}
\hline
\textbf{Dataset} & \textbf{Method} & \textbf{Acc} & \textbf{Prec} & \textbf{Rec} & \textbf{F1} & \textbf{AUC} \\
\hline
\multirow{2}{*}{rk\_dike}
& Nataraj & \textbf{0.9874} & \textbf{0.9697} & 0.9589 & \textbf{0.9642} & \textbf{0.9694} \\
& HilEnT  & 0.9809 & 0.9363 & \textbf{0.9599} & 0.9477 & 0.9641 \\
\hline

\multirow{2}{*}{rk\_lester}
& Nataraj & 0.9238 & 0.9238 & 0.9238 & 0.9238 & 0.9338 \\
& HilEnT  & \textbf{0.9355} & \textbf{0.9357} & \textbf{0.9354} & \textbf{0.9355} & \textbf{0.9397} \\
\hline

\multirow{2}{*}{rk\_ms}
& Nataraj & 0.9851 & 0.9850 & 0.9852 & 0.9851 & 0.9880 \\
& HilEnT  & \textbf{0.9954} & \textbf{0.9955} & \textbf{0.9954} & \textbf{0.9954} & \textbf{0.9971} \\
\hline

\multirow{2}{*}{rk\_self}
& Nataraj & 0.9811 & 0.9787 & 0.9773 & 0.9780 & 0.9820 \\
& HilEnT  & \textbf{0.9863} & \textbf{0.9846} & \textbf{0.9834} & \textbf{0.9840} & \textbf{0.9850} \\
\hline

\end{tabular}
\end{table*}
Furthermore, we test our image methods efficacy using a state of the art classification pipeline \citep{andriani2025cnnautomic}. Table~\ref{table:cnnautomic_comparison} reports the performance of the CNN-AutoMIC architecture when trained on images generated using the Nataraj grayscale visualization and our HilEnT representation. As expected, the two visualization strategies remain largely comparable across most metrics and datasets, reflecting the robustness of CNN-AutoMIC to different input encodings. Notably, HilEnT achieves consistently stronger performance on the Lester, Microsoft, and self-collected datasets, showing improvements across all five metrics. For the Dike dataset, Nataraj remains marginally higher in accuracy, precision, F1, and AUC, while HilEnT produces a slightly higher recall, suggesting that the Hilbert-entropy representation emphasizes different structural cues within the malware samples. Importantly, these accuracy differences are relatively small, while HilEnT offers significant advantages in inference speed (Table~\ref{table:runtime_comparison}), achieving the lowest per-sample latency among all evaluated configurations. This indicates that HilEnT provides a favourable trade-off: comparable or improved classification performance for most datasets while substantially reducing computational cost, making it particularly suitable for large-scale or real-time malware detection pipelines.

\subsection{Limitations and Future Work}

While HilEnT achieves competitive performance across multiple datasets, several limitations remain that motivate future work.

First, our experiments focus on Windows PE binaries (and OLE files in the Dike dataset), with a 5MB size limit and the exclusion of packed samples in the self-collected dataset. In addition, benign files for the Microsoft BIG 2015 and self-collected experiments are drawn from a single external benign corpus. Future work will extend HilEnT to other executable and document formats (e.g. ELF, Android, office and PDF malware), relax strict size and packing constraints, and evaluate on more diverse benign corpora that better reflect realistic deployment environments.

Second, HilEnT currently relies on a single static modality: raw bytes transformed into images. This is an intentional design choice that avoids executing malware and enables low risk deployment on endpoints without sandboxing. However, it also means that other informative static artefacts (such as PE header fields, section level metadata, imported API profiles, or string features) are not yet exploited. A natural extension is to build multimodal static architectures. This would combine HilEnT images with complementary static representations, which would allow us to study how safely collected dynamic telemetry could be fused with HilEnT without changing the static only nature of the endpoint detector.

Finally, we have not systematically analysed the robustness of HilEnT to subtle code manipulation. In particular, the response of our methodology to very slight malware injection into benign files. This may produce overlapping HilEnT images for infected goodware, clean benign, and fully malicious samples. Future work will investigate the sensitivity of HilEnT to low footprint injections, padding and other minor modifications that preserve functionality. We also hope to study mechanisms such as multi-scale entropy, local anomaly scoring, and uncertainty estimation to better separate these overlapping regimes and improve robustness in adversarial or grey area scenarios.

\section{Conclusion}\label{sec:conclusion}
Malware detection and classification are critical, with the high number of new malware released which target varying industries. For malware visualization, we proposed a novel malware binary to image transformation technique \textit{HilEnT} based on a combination of Hilbert curve transformation, benign and malware entropy cutoff comparisons to obtain a three channel image. We further utilized this image to perform malware detection using three different application driven algorithms: simple CNN based supervised learning, HOG-PCA enhanced supervised learning for time-performance consideration and few-shot learning based approach for practical cases with limited samples. We evaluated the performance on four datasets to achieve comparable malware classification performance to the state-of-the-art methods with a focus on reduced inference time. 

\backmatter

\section*{Abbreviations}
\begin{description}
	\item[CNN] Convolutional Neural Network
	\item[SVM] Support Vector Machine
	\item[MLP] Multilayer Perceptron
	\item[CSNN] Convolutional Siamese Neural Network
	\item[HOG] Histogram of Oriented Gradients
	\item[PCA] Principal Component Analysis
	\item[PE] Portable Executable
	\item[RGB] Red-Green-Blue 
	\item[FSL] Few-shot Learning
\end{description}

\section*{Declarations}

\begin{itemize}
	
\item Availability of Data and Materials

Publicly available datasets are referenced in the paper.

\item Funding

Not applicable

\item Acknowledgments

 We would like to express our sincere gratitude to the reviewers. 

\end{itemize}

\bibliography{sn-bibliography}

\end{document}